\documentclass[manuscript]{aastex}
\usepackage{outlines}
\usepackage{enumitem}

\slugcomment{to appear in The Astronomical Journal}

\shorttitle{Kepler Starspots}
\shortauthors{Basri \& Nguyen}

\begin{document}


\title{Double-Dipping: A New Relation between Stellar Rotation and Starspot Activity}


\author{Gibor Basri and Hieu T. Nguyen}
\affil{Astronomy Department, University of California,
    Berkeley, CA 94720}

\email{gbbasri@berkeley.edu} 





\begin{abstract}

We report the discovery of a new relationship between a simple morphological characteristic of light curves produced by starspots and stellar rotation periods. The characteristic we examine is whether the light curve exhibits one dip or two during a single rotation. We analyze thousands of Kepler light curves of main sequence stars from 3200-6200K. Almost all the stars exhibit segments of their light curve that contain either single or double dip segments (very few have more than two significant dips per rotation). We define a variable, the ``single/double ratio" (SDR) that expresses the ratio of the time spent in single mode to the time spent in double mode. Unexpectedly, there is a strong relationship between the SDR and the stellar rotation period, in the sense that longer periods come with a larger fraction of double segments. Even more unexpectedly, the slopes of the SDR-Period relations are a clear function of stellar temperature. We also show that the relationships of spot variability amplitude ($R_{var}$) to rotation period have similar levels of scatter, slopes, and dependence on temperature as the SDR-Period relations. Finally, the median $R_{var}$ of single segments tends to be about twice that of double segments in a given light curve. We offer some tentative interpretations of these new results in terms of starspot coverage and lifetimes. It will be fruitful to look further into this novel ``rotation-activity" relation, and better understand what information these aspects of light curve morphology bring to our knowledge of stellar magnetic activity. 

\end{abstract}





\section{Introduction\label{sec:Introduction}}
Sunspots have fascinated scientists since being documented by Galileo in the early 17th century. There have been centuries of observations of their behavior on the Sun, yielding insights into its surface and internal rotation and the operation of its magnetic dynamo. Photometric changes in other stars ascribed to starspots have also been observed for more than a century, but with vastly poorer time coverage and precision and little information on their sizes and positions. Things got slightly better with the advent of Doppler Imaging \citep{Vogt83} and more recently Zeeman Doppler Imaging \citep{Sem89}, which began to provide information on the size and position of spots (e.g. \citet{Stras09, See17}). 

With the advent of precision space photometry, particularly from the COROT \citep{Bag03} and Kepler \citep{Bor10} space telescopes, this situation has been markedly improved. In particular, the Kepler mission provides nearly continuous coverage over 4 years for well over a hundred thousand stars, with sufficient precision to detect individual sunspots if the Sun were hundreds of parsecs away. This has yielded a wealth of information on stellar rotation periods (e.g. \citet{McQ14}), and has been mined to learn more about spot coverage and variability, and potentially differential rotation for thousands of stars. There have been many other detailed analyses of individual stars with various types of mapping techniques employed, but here we are concerned with statistical behavior of groups of stars. The work that is closest in spirit to this paper is an analysis of about a thousand Kepler stars by \citet{Ark18}. Although their sample is much smaller and restricted to the lower half of our period distribution, they also consider the behavior of the stars per rotation period, and make some use of period harmonics. Another paper with a restricted sample size (but only 2 discrete periods) is that of \citet{Gil17}. They concentrate on the question of spot lifetimes, and utilize characteristics of the entire light curve, but are also affected by some of the issues of concern here. This paper takes a much simpler approach (which is much more easily visualized) and considers a much larger sample with the full period range. Some important issues remain for all work to date on how to translate Kepler pipeline differential intensities into correct understandings of starspot distributions and evolution, and the actual information content of broadband light curves (particularly a better understanding of the degeneracies they hide).  

Starspots are, of course, one manifestation of stellar magnetic activity.  One of the most basic facts about this activity is that it depends on stellar mass, age, and rotation period. A variety of ``rotation-activity" relations have been studied over the past several decades (e.g. \citet{Noys84, AR14}). They have concentrated on diagnostics of magnetic heating, from the chromosphere to the corona. In general, the faster a star with a convective envelope and a given mass is rotating, the stronger is the magnetic activity and the brighter its emission diagnostics. It is also true that this activity induces magnetic braking in most cases, which causes stars to spin down over time and thus their activity to decrease. This leads to the method of gyrochronology, which relates age, mass, and rotation (e.g. \citet{Meib15}). Much less work has been done on how starspots fit into this picture because of the lack of appropriate data until recently. General relations between the amount of starspot variability and stellar rotation have been shown, for example, by \citet{TR13, McQ14, Gil17, Ark18}. This paper looks into a barely explored aspect of starspot signatures, and develops a new way of gaining information based on the changing shapes of stellar light curves over time.

    \subsection{Kepler light curve morphologies\label{sec:morphology}}
    
In examining Kepler light curves, one of the most obvious features is that a given light curve often contains segments with systematic changes of amplitude over time. One can determine the rotation period of a star if it has an asymmetric spot distribution that retains some features for several rotations and produces a measurable signal. It is then possible to characterize properties of the light curve over each rotation. One of the simplest of these is a count of how many dips occur during a rotation. It turns out that because of the poor spatial resolution inherent in the light curve by itself, there are typically only one or two dips per rotation. It is important to keep in mind that although it is tempting to associate a dip in the light curve with a spot of a given size, the reality is that it reflects the sum of spots over most of a hemisphere. Thus, when we refer to a ``spot", we are generally talking about a number of spots that have a spatial correlation with each other (which might be strong or fairly vague). Nonetheless, we will continue to use the term ``spot" for convenience.

For two long-lived spots of equal size rotating with different periods, the light curve will have a larger amplitude when they are relatively near each other on the stellar surface (generating a single dip while in view during a rotation) but a smaller amplitude with double dips per rotation when they are on opposite sides of the star and not reinforcing each other. The advantage of integrating over a whole rotation period is that (unless spots grow and decay on the timescale of a rotation) all potentially visible spots will produce a signal for part of the rotation, regardless of how they may be distributed. 

Another possibility is that the changes in amplitude of the light curve are due to spot evolution, with larger light deficits being produced when there are more or larger spots present. One would have to also invoke preferentially active longitude bands or long spot lifetimes (and no differential rotation) if these spots are to preferentially produce single or double dips over many rotations, since the hemispheric distribution would have to be maintained at some level. It is less obvious in this case what the relation between single and double dip amplitudes should be. In practice, it will often be the case that both differential rotation and spot evolution are important. For this paper it does not matter what is producing the changes in the light curve; we are examining the morphologies themselves.

\section{A New Diagnostic: the Single/Double Ratio \label{sec:SDR}}

One of the most noticeable characteristics of Kepler light curves of main sequence stars is that many of them show obvious periodicity (when they show measurable variability at all). This may not seem surprising until one remembers that the Sun itself does not usually present an obviously periodic light curve. The methods used to find stellar rotation periods on Kepler stars fail in significant chunks of the solar light curve (despite its excellent signal-to-noise). However a number of authors have been able to determine rotation periods for tens of thousands of stars in the Kepler sample. An oft-quoted collection of them is in \citet{McQ14}, and we utilize that paper both as a source of targets and for adopted rotation periods. It is worth pointing out here that there is an even larger body of main sequence stars which do not display a measurable rotation signal. These are likely to be older stars with too little activity, so the conclusions of this paper really only apply to stars that are above that threshold of activity.

\citet{McQ14} used an auto-correlation method, and noted the common appearance of two repeating but dissimilar correlation function features which occur with a certain time spacing, and about half that spacing. This leads to the question as to which spacing represents the true stellar rotation period (the presence of two periods can also confuse periodogram methods of finding periods). \citet{McQ14} noted that generally one of the features looks weaker than the other, and that pattern often repeats for a while. They used this asymmetry to distinguish between the case of two features per period or just one, allowing an assignment of a true period (rather than its first harmonic). 

Obviously, it is hard to claim there are two dips per rotation unless there is some sort of repeating asymmetry between them; if they all looked the same or fluctuated randomly then the assigned period would likely be the spacing between adjacent dips. Another good way to distinguish single from double dips is to look for the largest repeating separation between dips (widest light curve features). That is because it is not really plausible for a star to require two rotations to produce what looks like a single dip; the star would have to get steadily darker over one rotation then steadily brighter in a similar fashion over the next rotation. This is especially unlikely if the single dips repeat several times. These points suggest further techniques for distinguishing between true and half periods by keeping track of the separations between dips as well as the amplitudes of successive dips.
 
The basic morphological question we ask is: what percentage of the time does a star display a single dip per rotation, and what percentage of the time are there double dips per rotation? It is this characteristic of the light curves that is the primary point of this paper. We specifically define the quantity of interest as the logarithm of the ratio of the time spent in single dip mode to the time spent in double dip mode (which we will refer to as the single/double ratio, or {\bf SDR}). This ratio is evaluated over the whole Kepler record for each star, typically a four year timespan. Thus an SDR of 0.0 means that the star spent as long in the single dip mode as in the double dip mode, an SDR of -1.0 means that it spent ten time longer in the double mode, and an SDR of 1.0 means it spent ten times longer in the single mode. The main results in this paper are that {\it the SDR is a strong function of the rotation period of the star, and the slope of the relation between SDR and rotation period is a clear function of the effective temperature} of the star. 

\begin{figure}
    \begin{center}
    \epsscale{1.0}
    \plotone{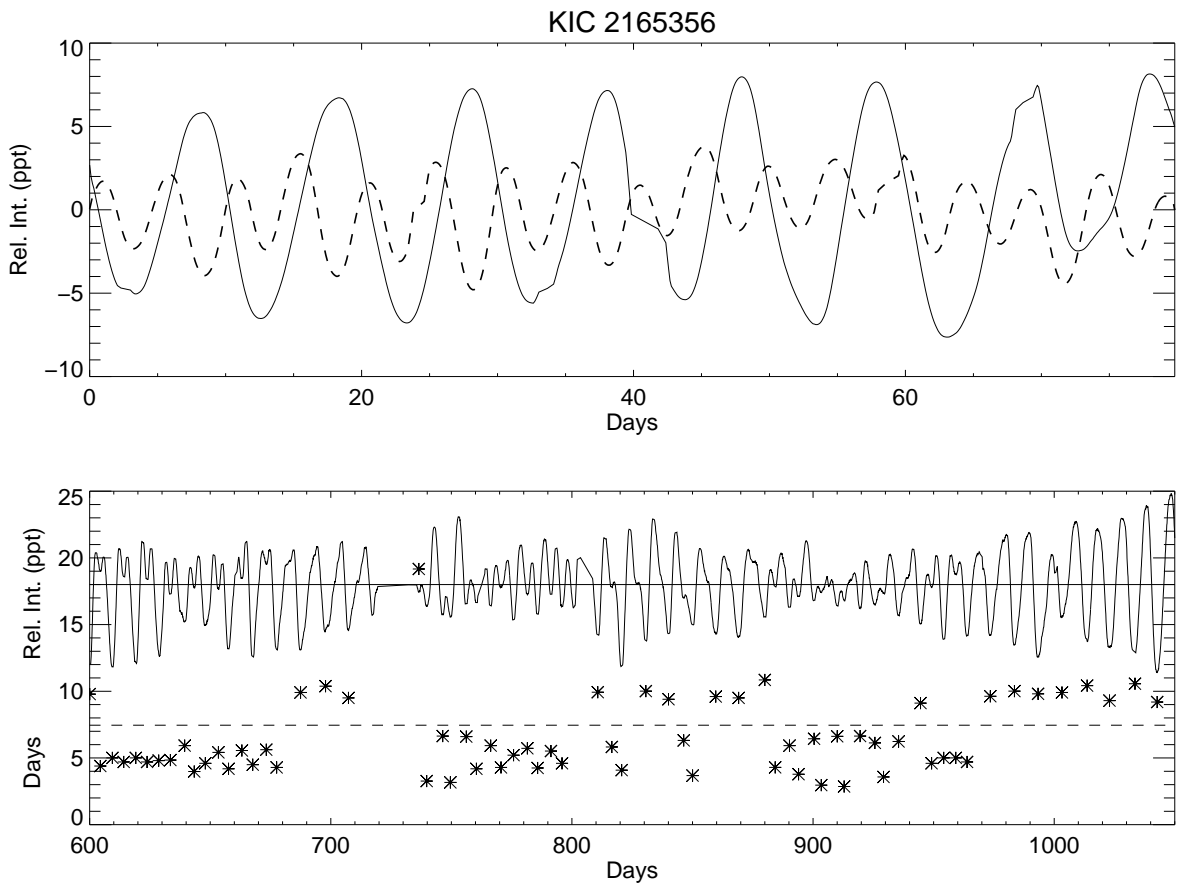}
    \caption{Portions of a particular star's light curve that contains both single and double dip segments. The top panel shows a sample single dip segment (solid) and double dip segment (dashed) on the same relative timescale. The actual starting days are 1030 and 480 respectively. The ordinate is the differential intensity in parts per thousand (ppt). The bottom panel's upper plot shows a longer section of the full light curve. For that the ordinate is also in differential ppt, but with zero displaced up to the solid line. The asterisks below mark the separation between successive dips; for those the ordinate is in days (with zero at the bottom). The dashed line indicates 75\% of the rotation period of 9.94 days, the boundary below which we classify dips as double. These lower points straddle half the period, and sometimes display coherent temporal structure.
    \label{fig:LCsegs}}
    \end{center}
\end{figure}        

The data used in this analysis are PDC-SAP pipeline products from 2015, obtained from the Kepler Archive at the MAST. We checked a number of cases from the last release of this pipeline product (DR25) and found it is not necessary for the purposes of this paper to re-download the whole dataset. The long cadence Kepler data have a 29 minute time resolution, which is far finer than needed to track the variability of spots. To speed things up and to eliminate features that are irrelevant to our purpose (because they are too fast to be produced by spots), we first rebinned the data by a factor of 10 to timesteps of about 0.2 days. To further reduce the effect of ``fake dips" due to noise instead of spots, we implemented boxcar smoothing in which the smoothing width for a given light curve is one eighth of the period found by \citet{McQ14}. This choice is heuristic; it means that the dips being counted are visibly significant and have timescales that are plausibly due to starspots. Our results are not qualitatively sensitive to the particular choice of smoothing. 

We locate the local maxima(peaks)/minima(dips) by comparing the relative flux at any given point in time with its neighbors, point by point throughout the whole light curve. We required that an extremum be present in a comparison of the nearest 4 points (to further eliminate too fast features). It turns out not to matter much whether one tracks dips or peaks, but we decided to use dips as our main focus since they are directly related to starspot presence. The question of small dips that are real but not counted here, and inflections that are not quite dips, will be taken up in a subsequent paper that looks at the behavior of these features in amplitude and phase over time (to address in detail the question of whether they are likely indicators of spot drifting).

After finding the time at which each dip occurs, we calculate all the temporal separations from one dip to the next. Then we compare these separations to the rotation period, to sort out the single dips and double dips. If the separations are greater than 75\% of the rotation period, the segments of light curve containing them are defined to be single dip segments; otherwise they are considered double dip segments. Figure \ref{fig:LCsegs} shows an example of what we are measuring. This star switches between single and double dip modes, often spending a few rotations in each (though not always). The single separations are mostly near the rotation period of 9.94 days; the double separations are more scattered around half that period. Structures that might be associated with phase drift are seen in the behavior of the some of the double dip separations (near 770 or 910). These tend to switch high and low as a secondary dip is closer to the previous and further from the next primary dip.   

\begin{figure}
    \begin{center}
    \epsscale{1.0}
    \plotone{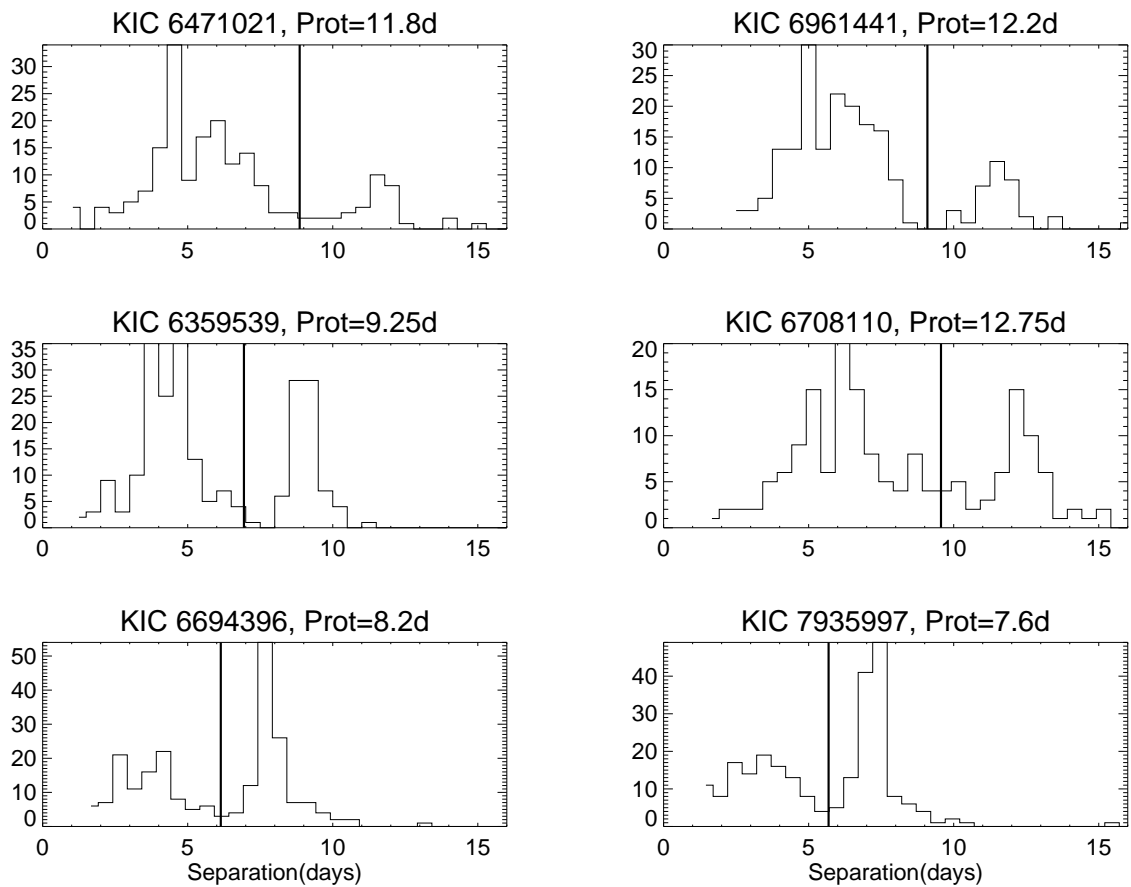}
    \caption{A few example histograms of dip separations. The top row has stars that are predominantly double dip, the middle row has about equal durations of single and double mode, and the bottom row has predominantly single dips. In each case the vertical line indicates 75\% of the rotation period; the point of division between single and double dips. Each case shows the minimum in the number of dip separations that occurs near this chosen boundary. Each title has the Kepler ID and the rotation period in days.
    \label{fig:SepHist}}
    \end{center}
\end{figure}

The choice of the boundary timescale midway between the rotation period and half the period is motivated in part by looking at histograms of the separations in light curves with significant segments of both single and double dips. One tends to see a concentration of separations around the period and half period, with a minimum about midway between them (Figure \ref{fig:SepHist}). Sometimes the separations approach our (somewhat arbitrary) boundary, but those are usually clearly associated with double dips. Of course, the histograms look different from star to star, and sometimes there is not so clear a separation between major histogram peaks (or even 2 major peaks at all). We are thus forcing a binary characterization of more complicated distributions, which should be kept in mind. However as we describe below, the stars behave in a very systematic way despite this simplification. 

Finally, we add up the durations of all the segments that have been characterized as either single or double to find the total duration that the star spends in single/double mode, and thereby determine the SDR. In a future paper we will analyze the behavior of the timing of dips (along with their amplitudes) in more detail, as this appears to be one fruitful approach to understanding the differences between differential rotation and evolution. 

\section{Analysis and Discussion}

\citet{McQ14} provide a well-determined and widely cited compilation of rotation periods for slightly over 34,000 stars from the Kepler target list. Most of these are main sequence stars; we restricted our sample by demanding that the log of the gravity obtained from the 2016 compilation of stellar parameters in the Exoplanet Archive \citep{Ake13} was greater than 4.3. This was also the source of our stellar temperatures. To avoid stars whose periods are due to pulsations rather than starspots, we restricted our period range to be greater than 3 days, and the temperatures to be between 3200-6200K. This produced an overall sample of 26,628 stars, which we then broke into 15 temperature groups of 200K each (Table 1). The groups do not have the same number of stars in them due to the way that the Kepler team chose targets; the coolest groups are particularly small and may not be fully statistically representative of their population. The period ranges for the cooler star samples are larger than for warmer stars because those samples contain more long period cases (e.g. \citet{McQ14}). In addition, we cut out the longest 2.5\% of periods in each group, because those longest period tails have much smaller densities of stars. Those tails all have flat SDRs with period (in the double regime), and the rotation periods are subject to more uncertainty because of potential problems with the pipeline reductions for such long timespans. 

\begin{figure}
    \begin{center}
    \epsscale{1.0}
    \plotone{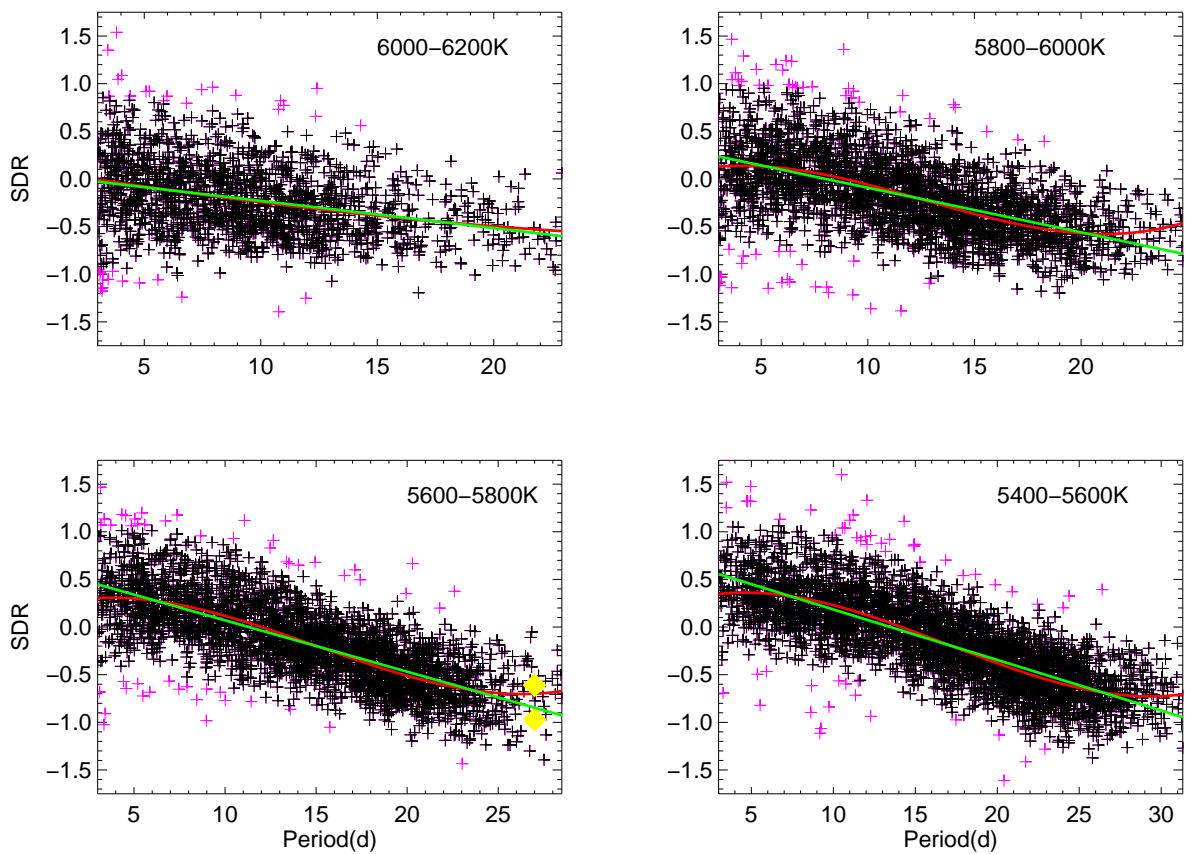}
    \caption{The relation between the SDR and rotation period in days for the warmest temperature groups. Note the changing period ranges in the different panels. A cubic fit to the sample is shown in red (note that SDR is a logarithmic quantity). A linear fit to each group of stars is also shown in green. Note that except for the warmest stars, the cubic fit tends to flatten a bit at either end. The purple points were not used for the final fits (they were outliers in initial cubic fits). The quiet and active Sun is indicated with the yellow diamonds.
    \label{fig:SDR1}}
    \end{center}
\end{figure}

\begin{figure}
    \begin{center}
    \epsscale{1.0}
    \plotone{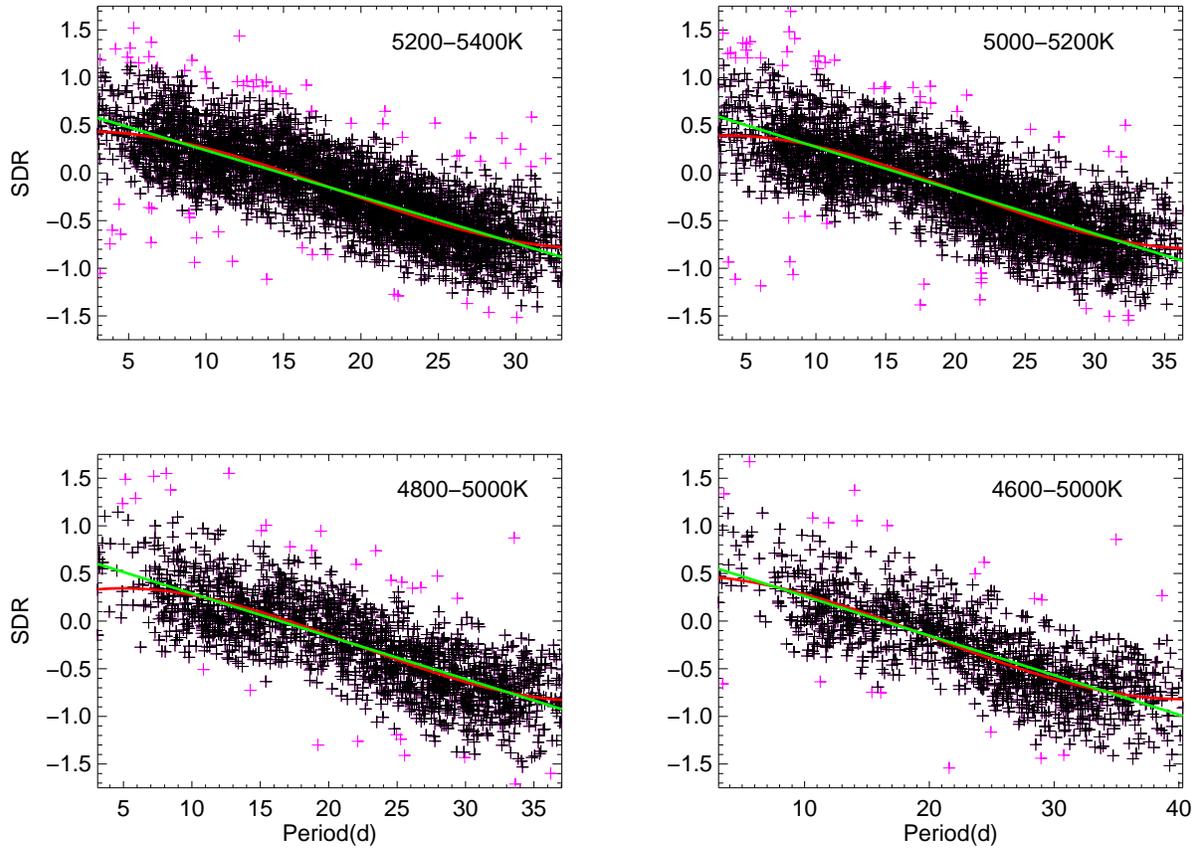}
    \caption{The same as Fig. \ref{fig:SDR1} but for the middle temperature groups. The cubic fits tend to flatten a bit at either end.
    \label{fig:SDR2}}
    \end{center}
\end{figure}

\begin{figure}
    \begin{center}
    \epsscale{1.0}
    \plotone{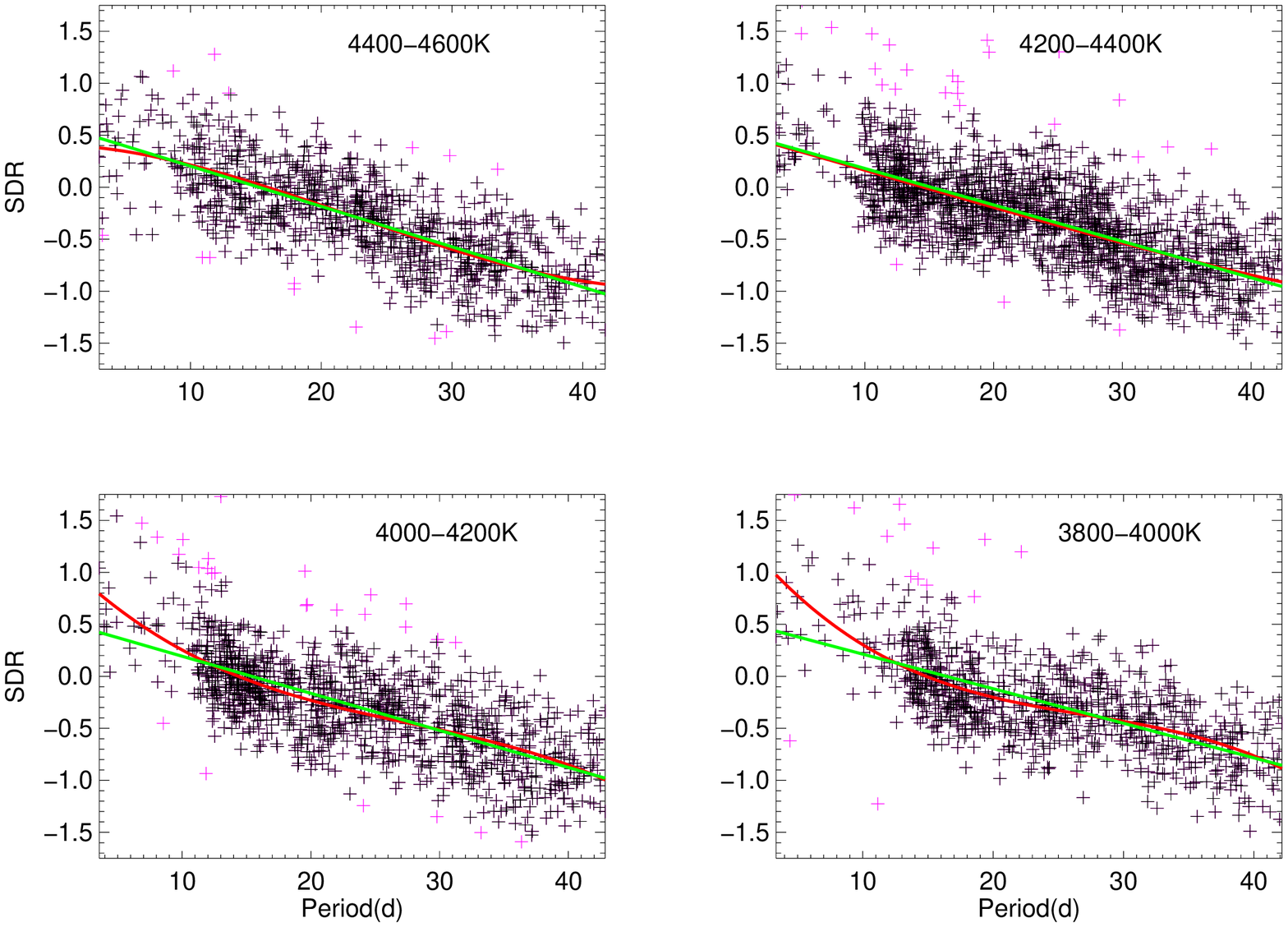}
    \caption{The same as Fig. \ref{fig:SDR1} but for the coolest temperature groups. The cubic fits for the two coolest groups rise at the short period end (unlike warmer groups).
    \label{fig:SDR3}}
    \end{center}
\end{figure}

Figures \ref{fig:SDR1} through \ref{fig:SDR3} display the main results of this paper. Each panel shows the relationship of SDR to rotation period for one of the temperature groups; we do not show the smaller groups below 3800K. Perhaps the first thing to note is that the points lie almost entirely between -1.0 and 1.0, which means that each of the light curves spends at least a tenth of the time in single dip mode and similarly in double dip mode. The points which have an SDR of 0.0 spend equal time in each state. Such cases are best for studying the ways in which light curves transition from single to double and back in detail (although we leave that for a later paper). It is also of note that the majority (between two-thirds and three-quarters in each group) of the stars have a negative SDR, which means that the double dip mode is preferred by most stars (column 3 of Table 1).

\begin{figure}
    \begin{center}
    \epsscale{1.0}
    \plotone{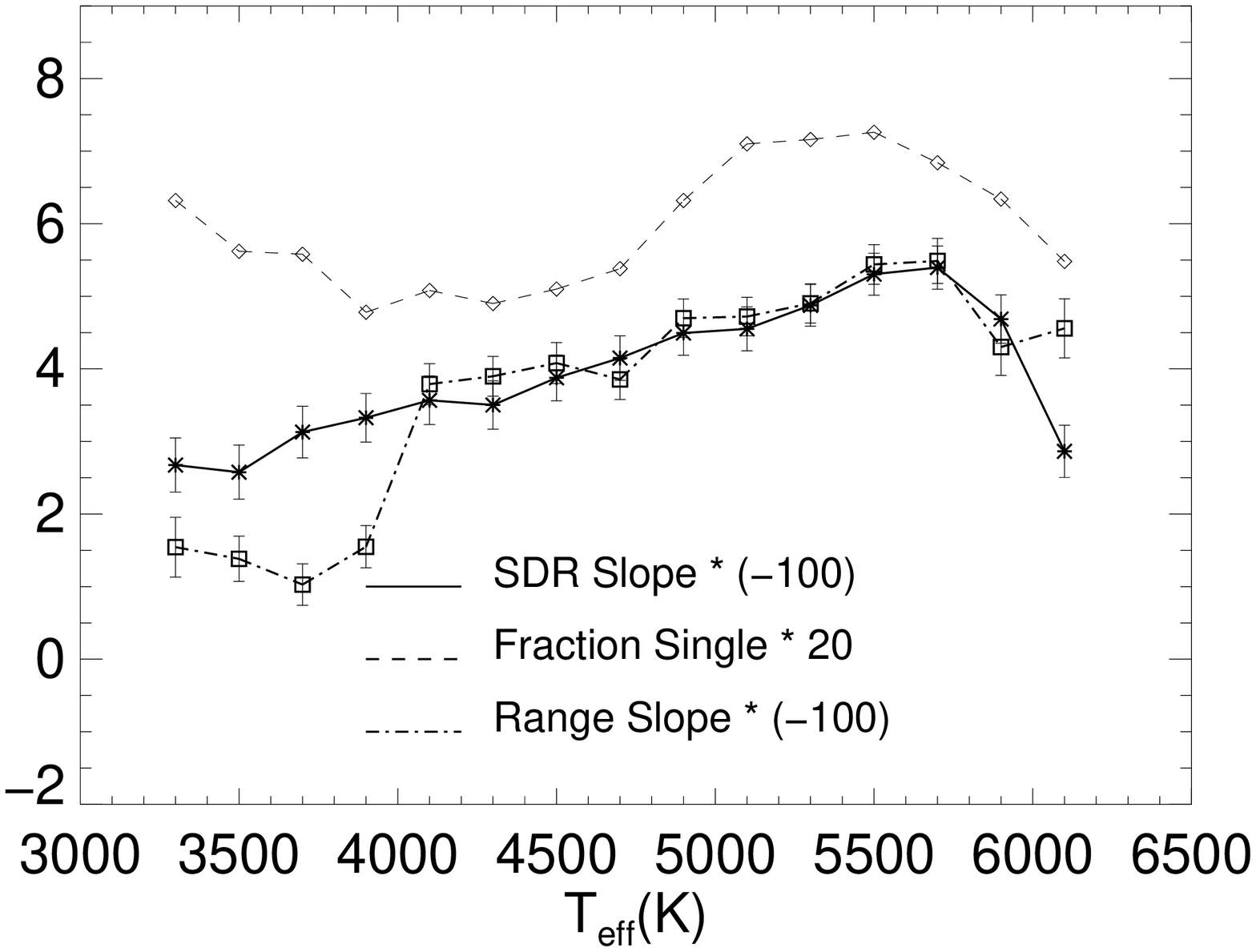}
    \caption{The systematic behavior of the slopes of the linear fits in Figs. \ref{fig:SDR1}  through \ref{fig:SDR3} with stellar temperature. Note that the negative of the slopes is shown; higher numbers indicate steeper slopes. Also shown is a representation of the fraction of the stars that have positive SDR, which is also systematic with temperature. For comparison, the slopes of the long period sides of the $R_{var}$-Period relations (discussed below) are also shown. They are quite similar to the slopes of the SDR-Period relations for most temperatures. We have adjusted the scales for each variable so they can be easily compared by eye in this Figure.
    \label{fig:SlpInt}}
    \end{center}
\end{figure}

Each panel shows a very obvious (inverse) correlation between SDR and rotation period, with the faster rotators tending to be more single and the slower rotators tending to be more double. We determined a simple linear fit between SDR (which is a logarithmic quantity) and period, and in each case it provides a good description of the trends. It is very interesting that both the slopes and the fraction of the points with positive SDR have systematic behaviors with stellar temperature, as shown in Fig. \ref{fig:SlpInt}. The slope steepens from 3200-5800K then starts to become shallower again for hotter stars. The steeper slopes also tend to go with a greater proportion of single dip light curves (except at the cool end). The values of the slopes and fractions are tabulated in Table 1. 

We also computed cubic fits to the points to capture potential curvature in the relations. This was done iteratively: after the first fit we discarded points with residuals of more than 2.5$\sigma$ before fitting again. There is a tendency for the middle temperature groups to show flattening at either end. The cubic fits provide comparability with the variability relations discussed next, which showed more obvious curvatures. It is clear for the SDR-Period relations, however, that linear fits are quite adequate.

\subsection{Spot Amplitudes}

In addition to computing the SDR for all the stars, we also examined photometric variability measures. There have been previous suggestions of a relationship between variability and rotation (e.g. \citet{TR13, McQ14}); the results here are consistent with previous analyses. Several variability diagnostics have been proposed and used with Kepler data. They can be defined on various time scales, for example, or with different methods of dealing with outliers. We employ the quarterly ``range" ($R_{var}$) that was used by \citet{Bas11} and others as our measure of variability. It is simply the difference in a quarterly light curve between its 5\% and 95\% levels (after sorting all points by relative intensity). Another popular choice is $S_{ph}$, which is the standard deviation of the intensity over a few rotation periods (e.g. \citet{Math14}). They produce the same qualitative results; $R_{var}$ has an advantage in general (though not here) that it does not depend on knowing the rotation period. We computed $R_{var}$ (in ppt) for each quarter for a star, then used the logarithm of the median quarterly $R_{var}$ value (for comparison with SDR, which is also a logarithmic quantity) as our final measure of variability for each star. The main difference with previous work is that we take a more detailed look at what is happening for different stellar temperatures. 

\begin{figure}
    \begin{center}
    \epsscale{1.0}
    \plotone{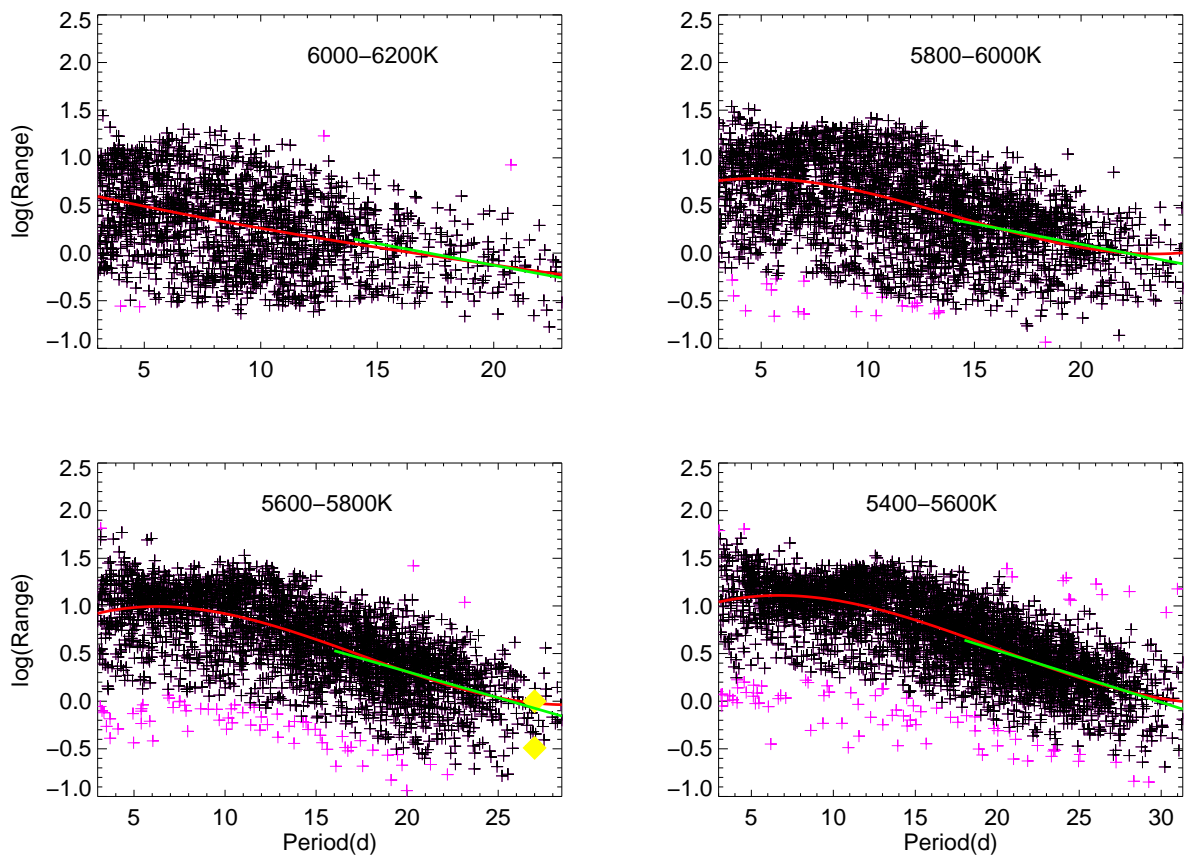}
    \caption{The relation between the rotation period (days) and the logarithm of the median quarterly $R_{var}$ (in ppt) for the warmest temperature groups. Note the changing period ranges in the different panels. An iterative cubic fit is shown in red (with 2.5$\sigma$ outliers shown in purple). There is a clear flattening at shorter periods for most temperature groups. A linear fit to the longer half of the period range is also shown in green for each group. The quiet and active Sun is indicated with the yellow diamonds.
    \label{fig:RngPer1}}
    \end{center}
\end{figure}

\begin{figure}
    \begin{center}
    \epsscale{1.0}
    \plotone{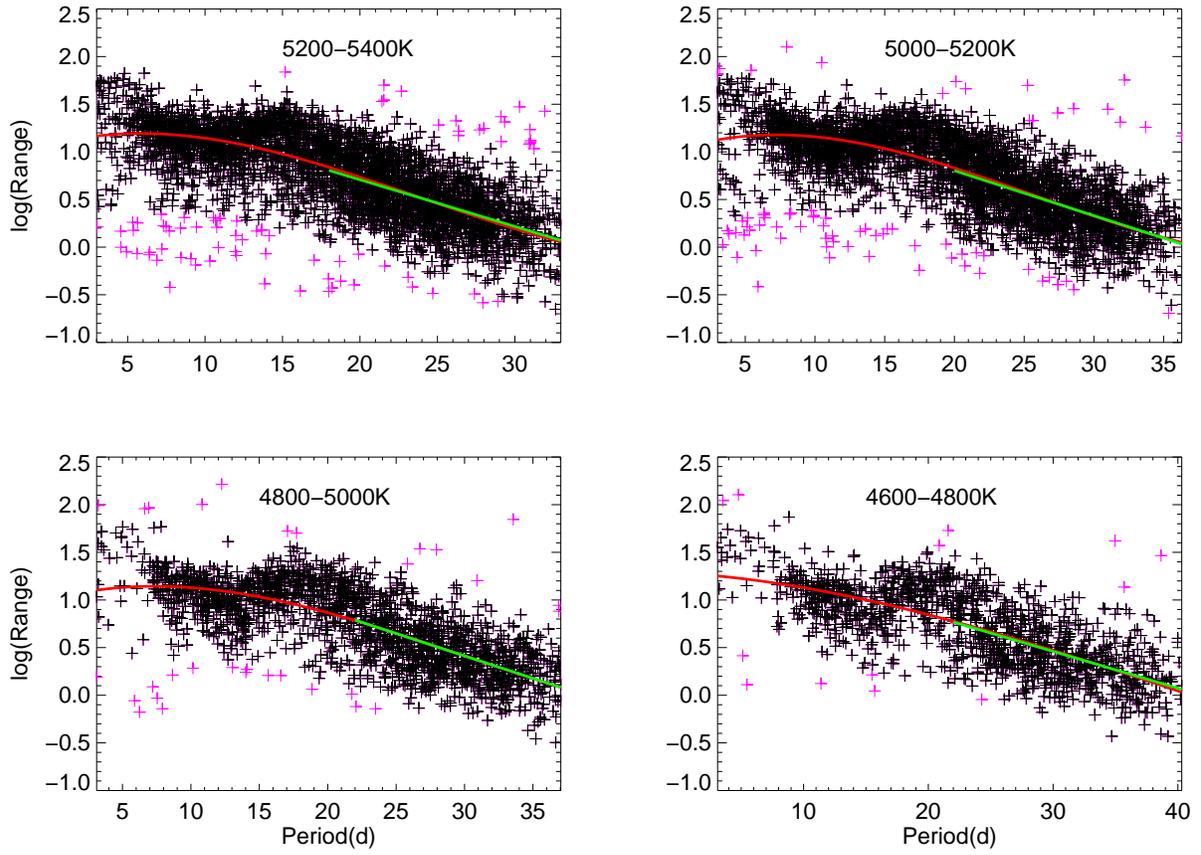}
    \caption{Same as Fig. \ref{fig:RngPer1} for the middle temperature groups.
    \label{fig:RngPer2}}
    \end{center}
\end{figure}

\begin{figure}
    \begin{center}
    \epsscale{1.0}
    \plotone{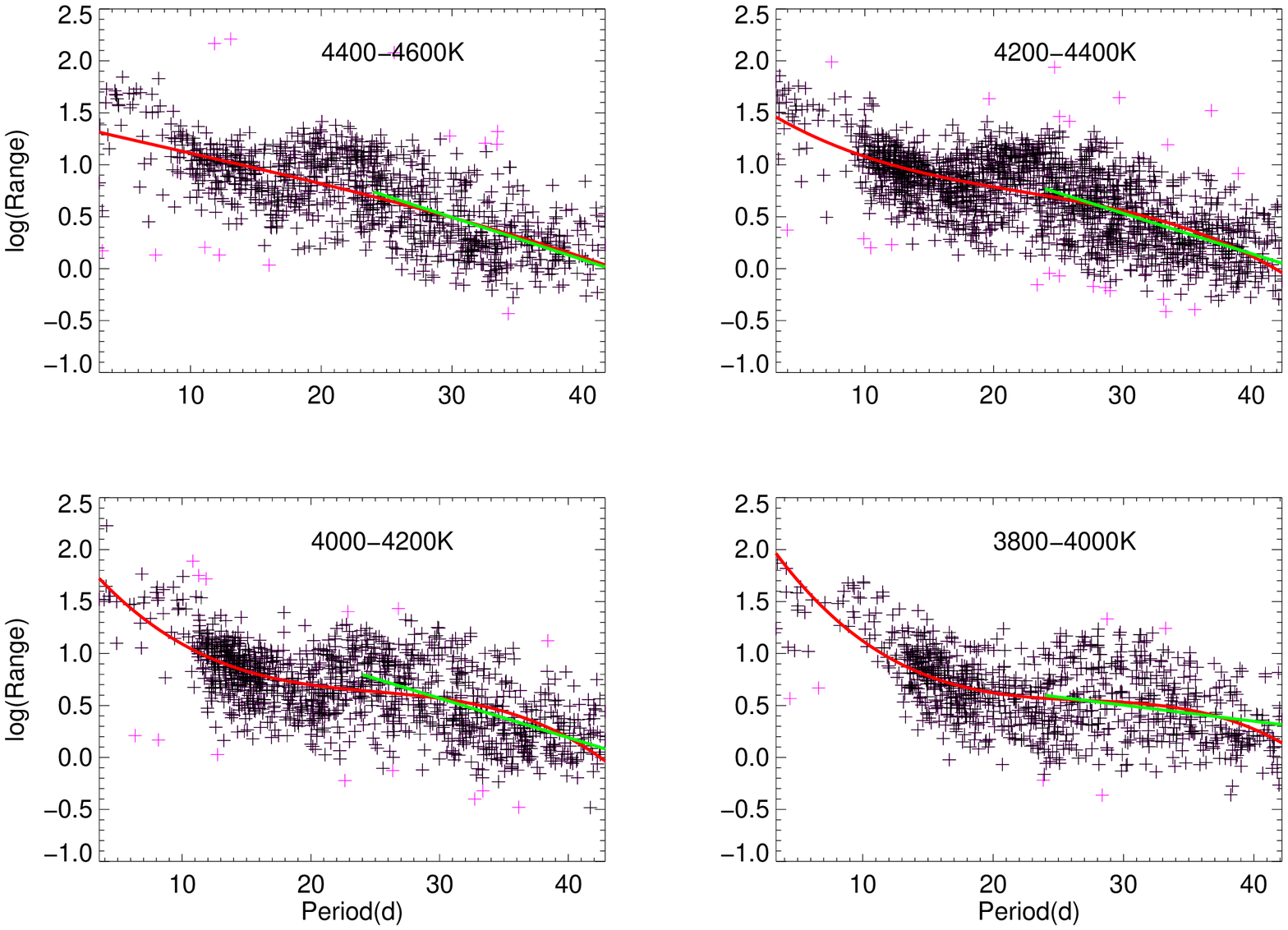}
    \caption{Same as Fig. \ref{fig:RngPer1} for the cooler temperature groups. As with the SDR, there is a switch to steeper upward slopes approaching short periods as one goes cooler.
    \label{fig:RngPer3}}
    \end{center}
\end{figure}

The $R_{var}$-Period relations are shown in Figs. \ref{fig:RngPer1} through \ref{fig:RngPer3}. For the warmest groups there is more scatter in the $R_{var}$-Period relationship, especially at shorter periods (they also span the shortest set of rotation periods). This is probably related to the beginnings of the tendency for stars with very shallow convective envelopes to remain more rapidly rotating due to the lack of magnetic braking, but they still exhibit some spot variability. 

As before we fit each group with a cubic polynomial in two iterations; in the second we discarded the points that were more than 2.5$\sigma$ away from the first fits. There is a tendency for the fits to flatten at rotation periods shorter than 15-20 days. The flattening moves longward for the cooler stars, which also extend to longer rotation periods. Between 4200-5800K the relation for the more rapid rotators resembles a ``saturation" behavior (very shallow slope), as is seen for a number of more traditional activity diagnostics, such as coronal X-rays (e.g. \citet{AR14}). For cooler stars (below 4200K) there is a group of more variable stars at periods less than 10 days, which give the short period power law a steeper slope (the opposite of what happens for the hotter stars), and the power law for these stars at longer periods is shallower than for the hotter stars. These trends continue down to 3200K. An odd dearth of lower values of $R_{var}$ appears just before the half-way point in period between temperatures of 4200-5200K. We are uncertain about the significance or meaning of that.

In order to compare with the linear fits for SDR-Period, we also produced linear fits for the longer period half of each temperature group. Note that the slopes of these longer period segments also show systematic behavior with stellar temperature (quite similar to the SDR slopes in Fig. \ref{fig:SlpInt}). This likely means that $R_{var}$ and SDR are related, as we will argue more explicitly below. 

We also evaluated how the scatter in the $R_{var}$-Period relations compares with the scatter in the SDR-Period relations. In Table 1 we tabulate the standard deviations ($\sigma_{SDR}$,$\sigma_{Rng}$) for both SDR and $R_{var}$ for each temperature group after subtracting the cubic fits versus rotation period. They are also shown as error bars in Fig. \ref{fig:SlpInt}. The scatter for the stars at the ends of the temperature ranges are highest, and the scatters for the $R_{var}$ relations are slightly lower than for the SDR relations in the middle temperature bins. Most of the relations have scatters of about a factor of two (0.3 dex), but there is not an important difference in the dispersion of the relations between SDR or $R_{var}$ versus period. This means that both the SDR and $R_{var}$ are ``rotation-activity" metrics, with similar levels of efficacy. The SDR has more to do with the geometry of the starspot distribution while $R_{var}$ has more to do with changes in starspot coverage, although these are not completely independent of each other.  

\begin{deluxetable}{rrrrrrr}
\tablewidth{0pt}
\tabletypesize{\scriptsize} %
\tablecaption{Temperature Groups of Stars and Fit Parameters.} 
\label{tab:TempGrps}
\tablehead{ 
\colhead{Temp. Group (K)} & \colhead{Sample Size} & \colhead{Frac.(single)\tablenotemark{a}} & \colhead{Slope (SDR) \tablenotemark{b}}  & \colhead{Slope ($R_{var}$) \tablenotemark{c}}  & \colhead{$\sigma_{SDR}$ \tablenotemark{d}} & \colhead{$\sigma_{Rng}$ \tablenotemark{e}} }
\startdata
3200-3400 &	158  &   0.316  &  -0.027  &  -0.015  &   0.374  &   0.412  \\
3400-3600 &	456  &   0.281  &  -0.026  &  -0.014  &   0.373  &   0.312  \\
3600-3800 &	673  &   0.279  &  -0.031  &  -0.010  &   0.356  &   0.286  \\
3800-4000 &	877  &   0.239  &  -0.033  &  -0.015  &   0.335  &   0.291  \\
4000-4200 &	1160 &   0.254  &  -0.036  &  -0.038  &   0.337  &   0.281  \\
4200-4400 &	1652 &   0.245  &  -0.035  &  -0.039  &   0.333  &   0.274  \\
4400-4600 &	983  &   0.255  &  -0.039  &  -0.041  &   0.316  &   0.282  \\
4600-4800 &	1232 &   0.269  &  -0.041  &  -0.039  &   0.307  &   0.275  \\
4800-5000 &	1807 &   0.316  &  -0.045  &  -0.047  &   0.308  &   0.264  \\
5000-5200 &	3364 &   0.355  &  -0.046  &  -0.047  &   0.303  &   0.265  \\
5200-5400 &	3650 &   0.358  &  -0.049  &  -0.049  &   0.287  &   0.271  \\
5400-5600 &	3532 &   0.363  &  -0.053  &  -0.054  &   0.289  &   0.272  \\
5600-5800 &	2877 &   0.342  &  -0.054  &  -0.055  &   0.298  &   0.310  \\
5800-6000 &	2595 &   0.317  &  -0.047  &  -0.043  &   0.333  &   0.390  \\
6000-6200 &	1613 &   0.274  &  -0.029  &  -0.046  &   0.360  &   0.407  \\
\enddata
\tablenotetext{a}{The fraction of light curves in this temperature bin that are predominantly single mode. }
\tablenotetext{b}{The slopes of the linear fits to SDR in Figs.\ref{fig:SDR1} through \ref{fig:SDR3}. }
\tablenotetext{c}{The slopes of the longer period linear fits to $R_{var}$ in Figs.\ref{fig:RngPer1} through \ref{fig:RngPer3}. }
\tablenotetext{d}{The 1$\sigma$ residuals for the cubic fits to SDR in Figs. \ref{fig:SDR1} through \ref{fig:SDR3}. }
\tablenotetext{e}{The 1$\sigma$ residuals for the cubic fits to $R_{var}$ in Figs.\ref{fig:RngPer1} through \ref{fig:RngPer3}. }
\end{deluxetable}

Finally, we examine the amplitude of variability ($R_{var}$) in the single dip segments compared with that in the double dip segments. Each ratio is computed by taking the median $R_{var}$ of all the single segments and the median $R_{var}$ of all the double segments in a given light curve. This is of interest in part because two different scenarios for why double dip segments appear could yield two different outcomes. If double dips arise primarily because of changes in hemispheric spot coverage, then one might expect $R_{var}$ to be somewhat similar in double dip and single dip segments. This could happen either through additional spot coverage on the hemisphere that was brighter during single dip times, or through reduction of the spot coverage on the darker hemisphere to levels more like the (formerly) brighter hemisphere. If, on the other hand, double dips arise because some of the spots on the darker hemisphere during single dip times migrate over to the brighter hemisphere (because of differential rotation), then the amplitudes of the double dip segments should be systematically smaller than for single dip segments. This would also yield the effect that the total spot coverage (integrated light deficit) per rotation would stay relatively constant across both single and double dip segments, whereas the first scenario would predict changing coverage. 

   \begin{figure}
    \begin{center}
    \epsscale{1.0}
    \plotone{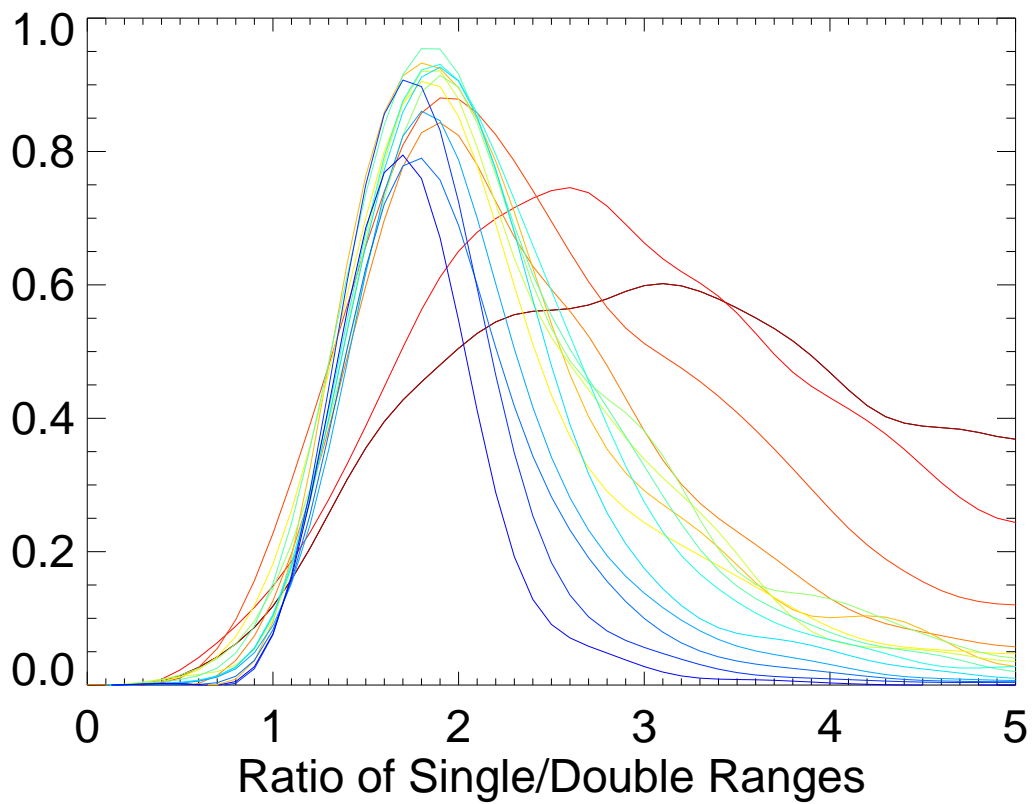}
    \caption{The ratios between the median $R_{var}$ of single segments to the median $R_{var}$ of double segments for all the temperature groups. The curves are smoothed versions of normalized histograms of the $R_{var}$ ratios for each temperature group. The colors vary between reddest for the coolest group to bluest for the warmest group. There is a strong tendency at most temperatures for the ratio of single to double $R_{var}$ to be nearly twice as high. More extreme ratios become increasingly common for cooler stars.
    \label{fig:SDrange}}
    \end{center}
\end{figure}        

Figure \ref{fig:SDrange} shows that there is a very clear preference for the single dip median $R_{var}$ to be nearly twice as large as the double dip median $R_{var}$. There are very few cases where they are equal or reversed. This result is consistent with the fact that predominantly double dip stars also have generally smaller median total $R_{var}$ (both SDR and median total $R_{var}$ decrease with increasing period). The fact that changing the mix of single/double segments will also change the mix of $R_{var}$ is probably the explanation for the correlated behavior of the slopes in Fig. \ref{fig:SlpInt}. There is a significant fraction of single/double $R_{var}$ ratios that are even higher than a factor of two for the coolest stars (higher contrast between single and double dips), and the distribution of ratios is also a function of stellar temperature as shown in Fig. \ref{fig:SDrange}. For stars above 5000K the single/double $R_{var}$ ratios are more tightly clustered around a factor of slightly under two. The tight clustering means that the ratio is maintained along the range of values of total $R_{var}$; it does not matter much whether the total variability is large or small. The broader distributions for the cooler groups are driven by a tendency for the single $R_{var}$ to scatter to larger relative values when the total variability is smaller, while the double $R_{var}$ tends to continue to track well with the total $R_{var}$.

The ratio of a factor of two between the single and double $R_{var}$ is roughly what would be expected if spots were simply redistributing themselves on the star to generate either single or double dips, with relatively constant coverage over a whole rotation (as argued above). Unfortunately the true physical situation is not yet clear. As we discuss below, the preponderance of higher amplitude single dips may instead (or in addition) be a result of a tendency for larger numbers of spots to more likely result in a single dip mode. 

It is also not currently clear how the light deficit signal per rotation (true total spot coverage) is actually changing, because there are still questions about how the Kepler reductions deal with absolute photometric changes. The pipeline differential light curves tend to have constant medians over each quarter, and if one supposes that the unspotted intensity lies at some constant value above the differential variations, that would imply relatively unchanging spot coverage (and that the variations are largely due to spot redistribution). But this may be an illusion due to the fact that real absolute variations on timescales of a few weeks or more are suppressed by the pipeline, in which case it is hard to know how the actual total light deficit due to spots is changing (not to mention the possible influence of faculae on this). We will study this question more in an upcoming paper that makes use of absolute calibrations by \citet{Mont17} for some Kepler stars. For now all we can say for sure is that the differential amplitudes of single dip segments tend to be about twice as large as those for double dip segments.

    \subsection {Possible interpretations}

One of the first questions these results raise is what sort of spot configurations lead to the single dip mode or the double dip mode in our light curves? For example, it is obvious that a configuration of 2 spots on opposite sides of the star will usually lead to a double dip. Less obvious is what minimum angular separation between the spots leads to a double dip. We utilized the analytic spot modeling described by \citet{Walk13} to begin to answer this. We started with $i=90^\circ$ models with 2 equatorial spots of equal size. In that case, the spots began to produce a double dip (as detected by our procedure) when separated by more than $105^\circ$. At a separation of $115^\circ$, however, the $i=30^\circ$ case is still classified as single, while higher inclinations are double. This is an effect of the decreased contrast at lower inclination, and shows that there are various complications in interpreting light curves. Changing the ratio of the spot sizes or their latitudes provides further complications. 

Before going further, it is appropriate to try to relate our results to those of two recent papers which have some similarities in what they investigate. The conclusions of \citet{Ark18} using about 1000 stars with the fast half of our period range seem to boil down to an observation that the size of the double dips (expressed through their half-period harmonic amplitude, somewhat akin to the behavior of our double dip $R_{var}$ over time) are less stable in time than the size of the single dips. We agree with that, but believe it is premature to tie that directly to the behavior of sub-surface diffusion and turbulence as they do. As we argue below, it is simply harder to produce a double dip morphology in the light curve (requiring more special spot configurations), and this naturally means that the double dips are more variable in their amplitude. Furthermore, differential rotation by itself can easily produce more rapidly changing double dips (in timing, depth, and amplitude) and less rapidly changing single dips. That is because the double dip amplitudes and positions are more sensitive to the actual spot configuration while the single dips only require a predominance of one hemisphere in coverage. As to the other conclusion of \citet{Ark18} that cool slow rotators show more stable (in amplitude) single dips than hotter faster rotators, we did not test this directly, but note that single dips (which we agree are more stable) are definitely more common in faster rotators at all temperatures.

\citet{Gil17} worked with a similar number of stars with similar temperatures, concentrated on periods of 10 and 20 days. They also reference double dip segments, referring to them as ``interpulse" features. Because they perform auto-correlations of a whole light curve, the signature of double dips is averaged in with single dips (and will be smaller with larger SDR). Their primary interest is in how the auto-correlation function decays as one looks a larger number of rotation periods out. Of course, if the star is switching from single to double mode (or vice versa) that will cause the auto-correlation function to decay faster. Given that the double dip separations are more spread than the single dip separations (Fig. \ref{fig:SepHist}), stars with lower SDR will also degrade the auto-correlation function more quickly. The primary conclusions of \citet{Gil17} are 1) the spot lifetime is longer for stars with greater variability, and 2) cooler stars tend to have longer spot lifetimes (which is in good agreement with \citet{Ark18}). In light of the our results, we could translate those to 1) stars with larger $R_{var}$ spend more time in single mode, and 2) cool stars tend to have larger $R_{var}$ and SDR at a given period. Both of these conditions will produce less degradation of the auto-correlation function. The first part of (2) was pointed out by \citet{Bas11} and reinforced by several later papers. The second part is partially true; the single fraction does increase for the coolest groups compared to stars of middle temperatures (Fig. \ref{fig:SlpInt}), although stars near solar temperature have even larger single fractions (although these are over the full period range). Perhaps it is the combination of the single fraction and $R_{var}$ that produces the cited result for the decay of the auto-correlation function for cool stars.

We do not mean to imply that the conclusions of \citet{Gil17} are wrong (indeed, we agree below that spot lifetimes may well be longer for higher SDR stars). We just want to point out that it is not yet clear whether spot lifetimes or spot configurations are being tested. Many authors have interpreted light curves in simple terms, imagining that the presence or absence of a secondary dip at about half a period reflects the behavior of a secondary spot on the other side of the star from a primary spot. We argue that this is overly simplistic, and can easily miss other different physical effects such as migration of some spots from one side to the other. In that case it would be a mistake to infer spot lifetimes from the behavior of the light curve. It is of interest to study in more detail how long each single/double mode lasts, and that would be one good thing to study next. Clearly stars that have an SDR of 0.5 or more spend most of their time in single mode, so their typical single segment must be longer than for lower SDR values.

To properly understand starspot light curve morphologies, in a following paper we will describe a far more complete exploration of the many-parameter space that spot models can occupy. Among the most relevant parameters are 1) basic spot characteristics like number, size distribution, and contrast, 2) the distributions in longitude and latitude: over what range?, are they random or grouped?, are they confined to stripes or belts?, 3) spot evolution: how often do spots appear? what are the growth and decay timescales?, 4) differential rotation: do they have different rotation periods? by how much?, how does it depend on their latitude?, and 5) the stellar inclination. Considerations such as these have not been carefully considered by most authors; one would like to know how they are manifested in metrics in intensity and frequency space. Metrics of the light curve morphology that should be studied include some measure of differential amplitude (like $R_{var}$), inferred spot coverage (which requires also knowing the unspotted ``continuum"), the SDR, periodogram structures (like double peaks), other measures of complexity (like auto-correlation functions and dip separation and depth histograms) and evolutionary behavior of the metrics. For each parameter set, we run thousands of trials with certain parameters randomized, and look at the statistical behavior of the metrics. Those results are far too lengthy to include in this paper, but we next present some preliminary information on how the SDR behaves.

One of the main conclusions from our extensive spot model trials is that the more common result in a general exploration of parameter space is a preference for positive SDR values. In other words, single dip light curve segments are decidedly more common than double dip segments for most of the configurations we have tried. This is in tension with the empirical results presented here (third column of Table 1), which show that most stars (other than the shortest period ones) tend to favor double dip morphologies. The only part of parameter space we have found so far that clearly favors double dips requires spot evolution with only a few spot groups present at a time and spot lifetimes not much over 1-2 rotation periods. As it happens, this describes the current state of the Sun (which indeed presents a strongly double dip light curve structure; Fig. \ref{fig:SDR1}). A sensible inference is that the strongly double dip stars (generally the longer period cases) have few spot groups, which probably don't last too long. Conversely, the fact that short period stars tend to have single dip structures and larger $R_{var}$ is consistent with the proposition that they have larger spot coverage (many spots present at the same time) and/or longer spot lifetimes. A remaining unsettled question is whether the spot coverage is relatively constant or changes substantially as a given star switches back and forth between single to double mode (resolving this requires a better understanding of how to interpret Kepler light curves, which we are also working on). 

Our tentative overall interpretation of our results is that the relation between SDR and rotation may be a diagnostic of a combination of spot coverage and lifetime, with both decreasing as the rotation period increases. Such a correlation is not surprising in a qualitative sense, but the combination of coverage and lifetime somehow manages to be closely related to the stellar rotation period. We do not yet understand why there are clear relationships between stellar surface temperature (mass) and the steepness of the SDR-Period relations. It is tempting to suppose that the Rossby number is somehow involved, since that is a widely used variable that introduces the stellar temperature (through the convective overturn time) into considerations of magnetic activity. We also are not currently certain we know why the single segment $R_{var}$ is commonly nearly twice the double segment $R_{var}$ within a light curve. These mysteries suggest that one might learn something about the stellar magnetic dynamo, or at least its production of starspots, from a better understanding of these relationships. Starspots may probe the connections between the depth (and perhaps vigor) of the convection zone and the geometry of field production. Spot lifetimes do indeed depend on diffusive processes that are related to sub-surface motions. That is the suggestion in the work of \citet{Ark18}, and although we are not currently convinced that their explanations have already been demonstrated, they are heading in the right direction. Clearly we need to better understand what the single/double nature of starspot light curves is really telling us. This paper is an early step in the analysis of that subject.

\acknowledgments

This paper includes data collected by the Kepler mission. Funding for the Kepler mission was provided by the NASA Science Mission directorate. Most of the data presented in this paper were obtained from the Mikulski Archive for Space Telescopes (MAST). STScI is operated by the Association of Universities for Research in Astronomy, Inc., under NASA contract NAS5-26555. This research has also made use of the NASA Exoplanet Archive, which is operated by the California Institute of Technology, under contract with the National Aeronautics and Space Administration under the Exoplanet Exploration Program. We would like to thank the referee for a careful reading of the manuscript and a number of suggestions that improved the paper.



{\it Facilities:} \facility{Kepler}, \facility{MAST}, \facility{Exoplanet Archive}.

\end{document}